\documentstyle[epsfig]{mn}

\begin{document}

%
%
   \title[Triggered star formation in expanding shells]{Triggered 
          star formation in expanding shells}

   \author[S. Ehlerov\'a and  
J. Palou\v s]{S. Ehlerov\'a\thanks{E-mail: sona@ig.cas.cz} and  
         J. Palou\v s\thanks{E-mail: palous@ig.cas.cz} \\
         Astronomical Institute, 
         Academy of Sciences of the Czech Republic,
         Bo\v cn\' \i \ II 1401, 141 31 Prague 4, Czech Republic}

   \date{Received 2001 August 6 / Accepted 2001 November 13}

   \maketitle

   \markboth{Ehlerov\'a \& Palou\v s: Triggered SF}{}

\begin{abstract}
We discuss fragmentation processes which induce star formation in dense walls 
of expanding shells. The influence of the energy input, the ISM scale-height 
and speed of sound in the ambient medium is tested. We formulate 
the condition 
for the gravitational fragmentation of expanding
shells: if the total surface density of the disc is higher than a
certain critical value, shells are unstable. The value
of the critical density depends on the energy of the shell and the
sound speed in the ISM.
   \end{abstract}


\section{Introduction} 

HI shells (or supershells) are structures identified in the distribution
of the neutral hydrogen (HI) in the Milky Way and in many nearby galaxies. 
The typical shell consists of a 
rarefied medium surrounded by a dense thin wall, in some cases 
expanding supersonically to the ambient interstellar medium (ISM). Dimensions
of shells lie in the range from a few 10 $pc$ to a few $kpc$. Energies
involved in their creation are of the order
of $10^{50} - 10^{54}erg$. Shells have miscellaneous shapes,
but many of them are nearly spherical, elliptical or cylindrical.
Observations of shells are reviewed in Brinks \& Walter (1998).

The energy needed for the creation of a shell originates either
in the combined effect of massive stars in an OB association 
(radiation, stellar winds and  supernova explosions), 
in the very energetic event connected to the gamma-ray burst 
(a hypernova or merging of compact companions in a binary system), or in
the infall of a high velocity cloud (HVC) to the galactic
HI disc. In this paper we consider the continuous energy input from 
massive stars of an OB association.

The density of the swept-up gas in walls of shells is higher than
the average density of the unperturbed ambient medium, which increases 
the probability of star formation.
Star formation in dense walls of HI shells is observed in some galaxies, 
e.g. in LMC (Kim et al., 1999), SMC (Stanimirovic et al., 1999), 
IC 2574 (Walter \& Brinks, 1999), Sex A (Van Dyk et al. 1998), etc. 
Its significance for the evolution of galaxies was studied by
Palou\v{s} et al. (1994) and others.

There are different mechanisms to trigger star formation: 
(1) a compression
of pre-existing clouds, (2) an accumulation of gas into a shell, 
(3) cloud collisions (Elmegreen, 1998), (4) shell collisions (Chernin et al.,
 1995). 
Here we focus on the mechanism (2) of the gravitational instability of mass
accumulated in a thin expanding shell. 
The growth of perturbations
was analyzed in the linear approximation by Elmegreen (1994), nonlinear 
terms have been included by W\" unsch \& Palou\v s (2001). 

Any density perturbation on the shell surface is stretched
by the expansion, while its
self-gravity supports its growth. The 
instantaneous maximum growth rate is
\begin{equation}
  \omega = -{3V \over R} + \sqrt{{V^2 \over R^2} +
  \left ({\pi G \Sigma_{sh}
  \over c_{sh}}\right)^2},
  \label{condx}
\end{equation}
where $R$ is the radius of the shell, $V$ is its expansion speed,
$\Sigma_{sh}$ is its column density  and $c_{sh}$ is the speed of sound 
within the wall. The  perturbation grows only if $\omega > 0$.

In galaxies with thin  discs shells differ from a sphere. 
To discuss  star formation triggered in expanding shells
without an \` a priori assumption on their shapes, 
we use 3-dimensional simulations. In a numerical code, the condition 
(\ref{condx}) is used and we quantify when and where the expanding 
shell starts to be unstable. 
This approach was first used in Ehlerov\' a et al. (1997, Paper I), 
here we extend parameter ranges and generalize results.  
We also discuss triggered star formation
observed in nearby galaxies and its connection to expanding shells.  

\section{The thin shell approximation} 

The energy input from an OB association 
creates a blastwave, which propagates into the ambient
medium (Ostriker \& McKee 1988; 
Bisnovatyi-Kogan \& Silich 1995). During the majority of the
evolution its radius is much larger than its thickness and the thin 
shell approximation can be used. In this approximation
the blastwave is considered to be an expanding infinitesimally thin 
layer surrounding the hot medium inside.

\subsection{The analytical solution}
  
The analytical solution of the expansion in the thin shell approximation 
was derived by Sedov (1959). 
In a static, homogeneous medium without the 
external or internal gravitational field, the blastwave
is always spherically symmetric, it sweeps the ambient matter
and decelerates. Neglecting the external pressure, and assuming the 
continuous energy input $L$, the self-similar solution for 
$R$, $V$ and $\Sigma_{sh}$ is (Castor et al., 1975):
\begin{eqnarray}
    R(t) & = & 53.1 \times 
    \left ({L \over 10^{51}erg Myr^{-1}}\right )^{1 \over 5} \nonumber \\
    & & \times
    \left ({\mu \over 1.3}{n \over cm^{-3}}\right )^{-{1 \over 5}}
    \times
    \left ({t\over Myr}\right )^{3\over 5} pc,
    \label{ssm1}
\end{eqnarray} 
\begin{eqnarray}
    V(t) & = & 31.2 \times 
    \left ({L \over 10^{51}erg Myr^{-1}}\right )^{1 \over 5} \nonumber \\
    & & \times
    \left ({\mu \over 1.3}{n \over cm^{-3}}\right )^{-{1 \over 5}}
    \times
    \left ({t\over Myr}\right )^{-{2\over 5}} km s^{-1},
    \label{ssm2}
\end{eqnarray}
\begin{eqnarray}
    \Sigma(t)_{sh} & = & 0.564 \times 
    \left ({L \over 10^{51}erg Myr^{-1}}\right )^{1 \over 5} \nonumber \\
    & & \times
    \left ({\mu \over 1.3}{n \over cm^{-3}}\right )^{4 \over 5}
    \times
    \left ({t\over Myr}\right )^{3\over 5} M_{\odot} pc^{-2},
    \label{ssm3}
\end{eqnarray}
where $n$ and $\mu$ are the volume
particle density of the ambient medium and the relative weight of one
particle (see also Paper I or Ehlerov\'a, 2000).

The formula (\ref{condx}) for the  
instantaneous maximum growth rate of perturbations shows that at early stages
of the evolution the shell is stable, as the fast expansion stretches
all perturbations which might appear. Only later, when $V$ is small,
$R$ large and $\Sigma_{sh}$ 
high, the self-gravity starts to play a role.
At the time $t_b$, when the growth rate $\omega $ becomes positive for the 
first time, the shell starts to be unstable and the fragmentation process  
begins. 

Inserting the analytical solution (\ref{ssm1}) - (\ref{ssm3}) 
for $R$, $V$ and $\Sigma_{sh}$
to the formula (\ref{condx}) we get
relations for the time $t_b$, 
the radius $R(t_b)$  and the expansion velocity 
$V(t_b)$:
\begin{eqnarray}
    t_b & = & 28.8 \times 
    \left ({c_{sh} \over kms^{-1}}\right )^{{5 \over 8}}
    \times
    \left ({L \over 10^{51}ergMyr^{-1}}\right )^{-{1 \over 8}}
    \nonumber \\
    & & \times
    \left ({\mu \over 1.3}{n \over cm^{-3}}\right )^{-{1\over 2}} Myr,
    \label{tb} 
\end{eqnarray}
\begin{eqnarray}
    R(t_b) & = & 399 \times
    \left ({c_{sh} \over kms^{-1}}\right )^{3 \over 8}
    \times
    \left ({L \over 10^{51}ergMyr^{-1}}\right )^{1 \over 8}
    \nonumber \\
    & & \times
    \left ({\mu \over 1.3}{n \over cm^{-3}}\right )^{-{1\over 2}}
    pc,
    \label{Rtb}
\end{eqnarray}
\begin{eqnarray}
    V(t_b) & = & 8.13 \times
    \left ({c_{sh} \over kms^{-1}}\right )^{-{1 \over 4}}
    \nonumber \\
    & & \times
    \left ({L \over 10^{51}ergMyr^{-1}}\right )^{1 \over 4} kms^{-1}.
    \label{Vtb}
\end{eqnarray}
The radius $R(t_b)$ is a lower limit to the distance  
on which the fragmentation (and triggered star formation) may 
happen; the expansion velocity $V(t_b)$ is the expansion velocity at which the
fragmentation process begins. Since in the subsequent evolution the shell
further decelerates it gives the upper limit  
to the random component of the velocity of newly created 
clouds (or stars). The relation (\ref{Vtb}) shows that $V(t_b)$ does not
depend on the density of the ambient medium $n$. This may be one 
reason why new clouds are born in different environments with similar 
velocity dispersions. 
  
Relations (\ref{tb}), (\ref{Rtb}) and (\ref{Vtb}) are 
derived under the assumption of the supersonic motion, i.e. 
$v_{exp}(t) > c_{ext}$. 
If the shell becomes a sound wave before it starts to be unstable
(i.e. before the time $t_b$), it will always remain stable. 
Assuming that the expansion velocity $V$ at $t_b$ has to be greater than
or equal to the sound speed in the ambient medium $c_{ext}$,
we can derive from the equation (\ref{Vtb})
the critical luminosity of the energy source $L_{crit}$,
for which the expansion velocity $V$ at the time $t_b$
equals the sound speed $c_{ext}$:
\begin{equation}
  L_{crit} = \left ( {c_{ext} \over 8.13\ kms^{-1}} \right )^4 
             \left ({c_{sh} \over kms^{-1}} \right )
             10^{51} erg Myr^{-1}
  \label{luml2}
\end{equation}
If the 
energy input is greater than $L_{crit}$, the shell starts to fragment;
if the input is smaller, the shell is always stable.
$L_{crit}$ does  not depend on the density of the ambient medium
but it is a strong function of its speed of sound.

Relations (\ref{tb}), (\ref{Rtb}) and (\ref{Vtb}), 
which apply
to the static and homogeneous ambient medium, can be extended to 
spherically symmetric cases
with the density gradient. Theis et al. (1998) show, that the density
gradient in the surrounding medium has to be shallower than isothermal
for the self-gravity of the accumulated matter to overcome the stability
supported  by the expansion.

The analytical solution for shells expanding in the 
exponentially stratified ISM  
was derived  by Maciejewski and Cox (1999). 
It assumes that the shape of the shell can be approximated 
by a prolate
ellipsoid. This assumption applies only to small shells created by energies
corresponding to about one supernova in the special ISM distribution. 
Unfortunately, for supershells with
energies in the range $10^{52} - 10^{54}$ erg, the assumption of ellipticity
fails. 

The linear perturbation theory does not
estimate the evolution of the instability after the time $t_b$ very well.
A better estimate of the evolution at $t > t_b$ is provided by the 
analysis including nonlinear terms (W\"unsch \& Palou\v{s}, 2001). 

\begin{figure*}
 \vglue-3.5cm
 \epsfig{figure=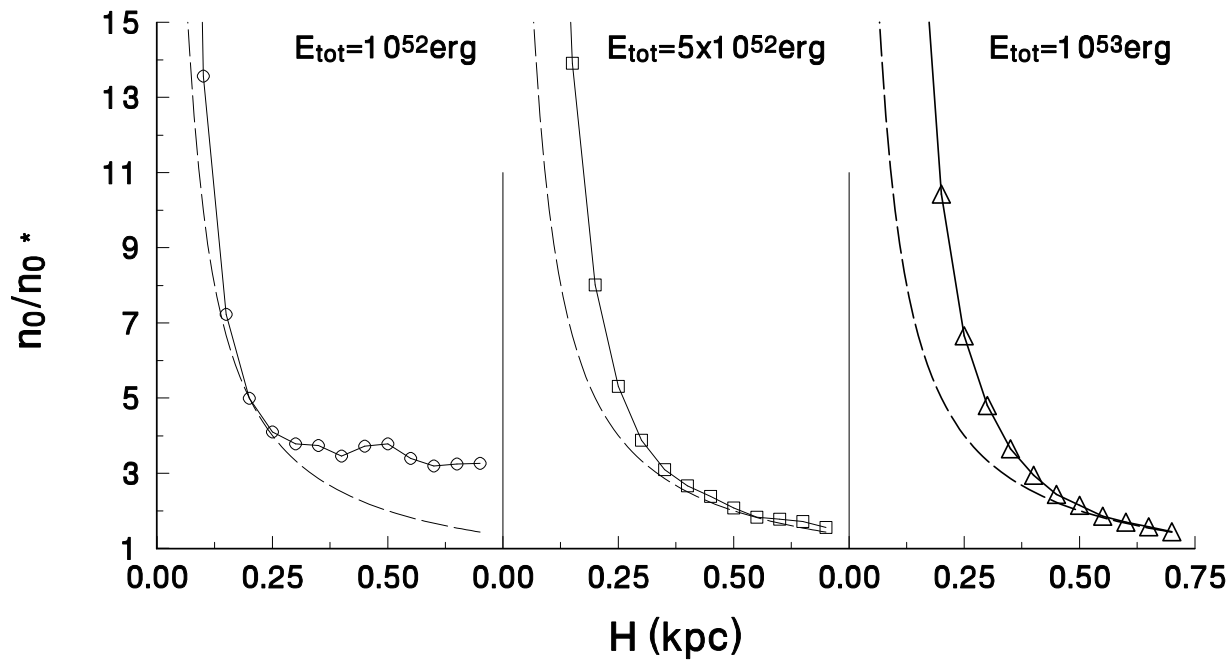,angle=270,width=18cm}
 \vskip3.5cm
 \caption{Fragmentation of shells in Gaussian discs, 
$c_{ext} = 5\ km s^{-1}$. The total input energy 
$E_{tot}$ is given in each panel.
The normalization density parameter $n_0^* = 0.265 \ cm^{-3}$ (left panel); 
$0.056 \ cm^{-3}$ (center); $0.031 \ cm^{-3}$ (right panel). 
Results of numerical simulations are shown by symbols connected by solid  
lines, they show the minimum density $n_0$ needed for the onset 
of the fragmentation. Dashed lines correspond to the constant gas 
surface density $\Sigma_{gas} = 14.5 \times 10^{20} \ cm^{-2}$ (left panel); 
$3.1 \times 10^{20} \ cm^{-2}$ (center); $1.7 \times 10^{20} \ cm^{-2}$ 
(right panel).}
 \label{fig1}
\end{figure*}

\subsection{Numerical simulations} 

The thin shell approximation has been applied in numerical simulations
in one and two dimensions by many authors (see Ikeuchi, 1998, for
a review on this subject). 
Models have been further extended into three dimensions by 
Palou\v s (1990) and Silich et al. (1996).

The code, which we use in this paper, is an improved and extended 
successor to the code of Palou\v s (1990), described in 
Paper I and in Efremov et al. (1999).
The thin shell is divided into a number of elements;
a system of equations of motion,  mass and  energy for each
element is solved and the fragmentation condition (\ref{condx})
is evaluated.

Numerical simulations include some effects, which are neglected in
the analytical solution. The most important are:
1) the pressure of the ambient medium and radiative cooling;
2) the finite, time-limited energy input from an OB association
(life-time of the source is $\tau $, we take $\tau = $ 15 Myr);
and perhaps the most obvious one 
3) the stratification of the ISM in galaxies.

Due to the disc-like ISM density distribution and  to the form of the 
gravitational potential, shapes of expanding shells are not spherical,
but elongated in the direction of
the density gradient; the differential rotation distorts shells
in the plane parallel to the galactic plane. 
In the case of higher input energies and thin discs, a blow-out may occur
and a fraction of the energy 
supplied by hot stars escapes to the galactic halo.
Blow-out decreases the influence of the 
energy source on the densest parts of the  shell in the galactic plane. 

Fragmentation properties
vary with the position on the shell. Typically, for shells
growing in a smooth distribution of gas with the $z$-gradient,
a dense ring is created in the region of the maximum density.
This is the most unstable part of the shell, while large lobes
in low-density regions are stable. In the following
we present results for the most unstable part of shells.
Energy sources, which are not located in the symmetry plane of
the disc, produce asymmetrical shells. However, as the fragmentation
takes place in the densest regions, it is not influenced very much
by the asymmetry of the shell. The blow-out effects (and the subsequent 
decrease of the ``effective'' energy) are enhanced in such a case.
 
The gravitational field of the galaxy may change the velocity field of 
the ISM and the shape of the shell, and thus conditions for the fragmentation.
Simulations show, that the differential
rotation has a smaller influence than the density stratification. 
Usually, two unstable regions appear on tips of an ellipse. 
Galaxies with little shear (e.g. dwarfs) do not influence the shells
even to that degree (see Ehlerov\'a, 2000).

The ISM in galaxies is not perfectly smooth but contains many small-scale 
density fluctuations (see Silich et al, 1996, for
the partial analysis of their influence), it is turbulent etc. 
However, coherent shells and bubbles are observed even in a rather 
turbulent ISM. Our numerical experiments in a medium with a hierarchy 
of density fluctuations show, 
that shells are mostly influenced by the large-scale gradients in the
ISM (with the possible exception of the very perturbed regions, where the 
coherent shock does not form).

\section{The critical surface density} 

To study the influence of the $z$-stratification,
we simulate shells in different types of the ISM distribution
with different disc thicknesses.  We fix the velocity dispersion in the
shell $c_{sh}$ (see eq. \ref{condx}) to a constant value: 
$c_{sh} = 1\  km s^{-1}$.
Results for the Gaussian stratification 
$n(z) = n_0 e^{-{z^2 \over H^2}}$
are given in Fig. \ref{fig1}.
Symbols denoting different $E_{tot}$ show the lowest density $n_0$ 
needed for the onset of the 
instability during the shell expansion. For a given disc thickness 
$H$, a total energy $E_{tot}$ ($E_{tot}$ is the total energy delivered
by the OB association to the shell: $E_{tot} = L \times \tau $) 
and a velocity dispersion $c_{ext}$, shells are unstable only if $n_0$ 
exceeds the depicted value. For lower values shells remain stable.
Dashed lines are lines of the constant gas surface density in the disc
($\Sigma _{gas}$).   

As can be seen in Fig. \ref{fig1}, lines of the constant disc surface 
density $\Sigma_{gas}$ approximately separate stable and unstable regions.
There are two types of deviations between the solid line separating stable
and unstable regions and $\Sigma_{gas} = const$ line:
\begin{itemize}
\item {\bf The blow-out effect.} It is visible in the middle and right panels
(i.e. in the case of higher energies) for thin discs: a higher density
$n_0$ is needed for the instability than predicted by $\Sigma_{gas} = const.$
The blow-out enables the leakage  of the energy to the galactic halo, leading
to the decrease of the effective energy and pressure pushing the densest parts
of the shell.
\item {\bf The small shell in the thick disc.} 
It is visible in the left 
panel (i.e. in the case of a low energy) and thick discs: 
a higher value of 
$n_0$ than predicted by $\Sigma _{gas} = const$ line is needed for the 
instability. Low energy shells are  generally small and in  thick discs 
they evolve in an almost homogeneous medium as they never
reach dimensions comparable to the thickness of the disc. Consequently, 
the value of the gas surface density of the disc is irrelevant. 
The critical value of $n_0$ has to be higher  than predicted by
$\Sigma_{gas} = const$ criterion, since a substantial
fraction of the ISM in the disc remains untouched by the shell.

A different behavior is observed in complementary 
simulations with $\tau = \infty $ (long-lasting energy supply). In this case  
shells fragment  at a radius $R(t_b) \simeq 1.5 H$. Inserting this condition 
into eq. (6) we get $n_0 \propto 1/H^2 $ for given $c_{sh}$ and $L$. The
deviation from this behavior in Fig. 1 is connected to the fact that 
$\tau < t_b$.  
\end{itemize} 

\begin{figure}
 \epsfig{figure=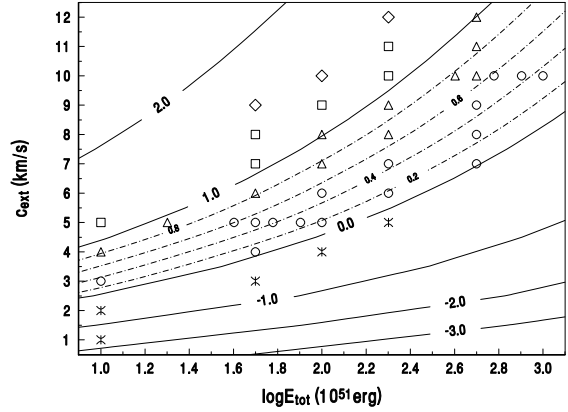,angle=270,width=\hsize}
 \vskip3.5cm
 \caption{The critical gas surface density $\Sigma_{crit}$ in the 
($log E_{tot}$, $c_{ext}$)  plane.
Various  symbols show $\Sigma_{crit}$ from numerical 
experiments: $* = \Sigma_{crit} \le 10^{20} cm^{-2}; \  
\circ  = 10^{20} < \Sigma_{crit} \le 4 \times 10^{20}; \    
\triangle = 4 \times 10^{20} < \Sigma_{crit} \le 10^{21};$ 
$\Box = 10^{21} < \Sigma_{crit} \le  2 \times 10^{21}; \  
\diamond = 2 \times 10^{21} < \Sigma_{crit}$. 
Contour lines are calculated by the formula (\ref{fit}) and  
labelled by the logarithm of $\Sigma_{crit}$ in $10^{20}\ cm^{-2}$.}
 \label{fig2}
\end{figure}

Other simulations with different disc profiles 
(Gaussian, exponential, multicomponent)
show that the critical
value of the gas surface density $\Sigma _{crit}$ 
does not depend on the profile.
$\Sigma_{crit}$ depends strongly on $c_{ext}$: 
for higher values of $c_{ext}$ the fragmentation
starts at higher values of $\Sigma_{gas}$. 
We derive the fit:
\begin{equation}
\Sigma_{crit}  =  0.27 \left ({ E_{tot} \over 10^{51} erg} 
\right )^{-1.1} \left ({c_{ext} \over km s^{-1}}\right )^{4.1} 
10^{20} cm^{-2}. 
\label{fit}
\end{equation}
This relation is illustrated in Fig. \ref{fig2}.

\section{Comparison to observations} 

In this paragraph we study one example of the HI hole with star formation 
on the rim. This is the structure number 35 in the galaxy IC 2574
(Walter \& Brinks, 1999). Taking the measured 
value of $c_{ext} = 7\ km \ s^{-1}$ and 
the total energy input from stellar winds and supernovae of the central
star cluster inside this shell, estimated by Stewart \& Walter (2000) as 
$E_{tot} = 4.1 \times  10^{52} erg$, and using  
the formula (\ref{fit}) we derive the critical gas surface density 
$\Sigma_{crit} = 1.3\times 10^{21}\ cm^{-2}$.  The gas surface density in 
the vicinity of this shell, $1.8 \times 10^{21}\ cm^{-2}$ 
(as given by Walter \& Brinks 1999), is higher than
the derived value $\Sigma_{crit}$, which is consistent with our hypothesis.

We may derive $R(t_b)$ using formula (6) for a homogeneous medium. 
For this hole we get $R(t_b) \sim 500$ pc, which is quite close to the observed
dimension.  The disc thickness $H$ of IC 2574 is 350 pc, which is less than
$R(t_b)$, and the assumption of homogeneity is not fulfilled.  In our opinion,
the condition (9) is more suitable for a discussion of the fragmentation
process since it takes into account the $z-$stratification.
      
A comparison of the HI distribution in IC 2574 to the $H_{\alpha}$ 
emission (which is very often connected to rims of HI shells) suggests,
that majority of $H_{\alpha}$ emission lies in the regions with $\Sigma_{gas} 
\ge 1.7 \times 10^{21}\ cm^{-2}$. If this is the critical gas surface 
density $\Sigma_{crit}$, we can estimate the value of the 
minimum energy
necessary to produce an unstable shell, which is  $E_{tot} = 3.8 
\times 10^{52} erg$.
It means, that in IC 2475 clusters with $\sim 40 $ massive 
stars and more can trigger star formation.

{\tiny
\begin{table}
 \centering
 \begin{tabular}{|l|l|l|l|}
 \hline
 Name  & $c_{ext}$ & $\Sigma_{crit}^{obs} $ & $E_{tot}$ \\
       & $ km s^{-1}$ & $M_{\odot } pc^{-2}$  & $10^{51}\ erg$ \\
 \hline
 IC 1613 & 7.5 & 3 & 257 \\
 Ho II   & 6.8 & 6 &  92 \\
 DDO 155 & 9.5 & 3 & 640 \\
  \hline
 \end{tabular}
 \caption{Values of the minimum input energy to trigger star formation
on the rim of the shell, $E_{tot}$, as derived from the observed
values of $c_{ext}$ and $\Sigma_{crit}^{obs}$.}    
 \label{tab1}
\end{table}
}
In Table \ref{tab1} we show the observed average
value of the velocity dispersion
$c_{ext}$ and $\Sigma_{crit}^{obs}$, the azimuthally averaged value of 
the gas surface density, 
below which star formation rate strongly declines, for three
galaxies selected from Hunter et al. (1998). $E_{tot}$ in the
last column of Table \ref{tab1}, derived from the  formula (\ref{fit}),
gives the value of the minimum input energy necessary to create an
unstable expanding shell. Corresponding numbers of massive stars in 
the star cluster differ for different galaxies: in
IC 2574 and Ho II less numerous clusters than in IC 1613 and DDO 155
are needed.

\section{Conclusions} 

We study  the gravitational instability of expanding shells
using the analytical solution and numerical simulations. The influence
of the total energy $E_{tot}$, the velocity dispersion of the ISM $c_{ext}$,
the  thickness
and the profile of the gaseous disc are analyzed with the following 
result: the critical value of the disc gas surface density 
$\Sigma_{crit}$ exists;
shells expanding in discs with the lower gas surface density are stable, 
shells
expanding in discs with the higher density are 
unstable, can fragment and form gaseous clouds and later stars.

Values of $\Sigma_{crit}$ for reasonable values of $E_{tot}$ and 
$c_{ext}$ are
of the order of $(10^{20} - 10^{21})\ cm^{-2}$, which coincides
with the value of observed threshold surface densities for  
star formation in galaxies (Kennicutt, 1997, 1998; Hunter et al., 1998).
This agreement may indicate, that the 
contribution of star formation induced by shells to total 
star formation is important.

We should distinguish 
between two cases: 1) the gas surface density $\Sigma_{gas}$ 
in the shell is increased 
compared to the average, which decreases the value of  
Toomre (1964) or Safronov (1960) parameter $Q$
for the local gravitational instability of the galactic disc; 
2) the material accumulated in the shell starts to be unstable according
to the formula (\ref{condx}). We label the first case 
an enhanced mechanism of spontaneous star 
formation, the second case describes triggered star formation. 
According to our simulations, the main
difference lies in the dependence on the velocity dispersion in the
unperturbed medium $c_{ext}$: the critical density $\Sigma_{crit} $ 
is in the case of spontaneous star formation directly proportional to 
$c_{ext}$, but in the case of triggered star formation it depends on
$c_{ext}^4$. A much sharper dependence of $\Sigma_{crit} $ on 
$c_{ext}$ is connected to the mass accumulation to the shell, which stops 
when the shell decelerates to $c_{ext}$.

The very steep dependence of  $\Sigma_{crit}$ on $c_{ext}$ indicates  
the importance of the 
self-regulating feedback for triggered star formation mode. 
Young stars in OB 
associations release the energy and compress the ambient ISM creating 
shells. In dense walls of shells, star formation may be triggered 
if the disc surface density surpasses a critical value 
$\Sigma_{crit}$. 
Star formation is accompanied by the heating of the ISM, i.e.
increasing $c_{ext}$. This leads to
the increase of $\Sigma_{crit}$ and a subsequent 
reduction of triggered star formation rate. 
The energy dissipation and cooling 
of the ISM decreases  $c_{ext}$ and $\Sigma_{crit}$ and 
increases  star formation, closing the self--regulating cycle 
of the triggered mode of star formation.

The heating of the ISM and the increase of $c_{ext}$ related to star 
formation  is a  localized process operating on scales of  
$100\ pc - 1\ kpc$ in size. Consequently, triggered star formation 
is  a more local process than  
the spontaneous mode. The less steep dependence of 
$\Sigma_{crit}$ on $c_{ext}$ for spontaneous star formation means, 
that this mode may be also effective in regions, where $c_{ext}$ is 
increased and the triggered mode of star formation temporarily stopped.      

\section*{acknowledgements}
We would like to thank B. G. Elmegreen,  L. S. Schulman and to anonymous 
referee for helpful comments.
Authors gratefully acknowledge a financial support by the Grant
Agency of the Academy of Sciences of the Czech Republic under the grant 
No. A3003705/1997 and
a support by the grant project of the Academy of Sciences of the
Czech Republic No. K1048102.

{}

\end{document}